\pdfoutput=1

\documentclass{article}
\usepackage{graphicx}
\usepackage{tikz}
\usepackage{standalone}
\usepackage{pgfplots}
\usepackage{amsmath,amssymb}
\usepackage{siunitx}
\usepackage{paralist}
\usetikzlibrary{positioning, calc, chains,shapes.geometric, shadows, shapes.misc, fit}

\usepackage{amsmath}
\usepackage{amssymb}
\usepackage{bbm}
\usepackage{physics}
\usepackage{siunitx}
\usepackage{trfsigns}
\usepackage{pifont}
\usepackage{balance}

\DeclareMathOperator{\ExpOp}{\mathbb{E}}

\usepackage{pgfplotstable}
\usepgfplotslibrary{groupplots}
\usepackage[english,capitalize,noabbrev]{cleveref}
\usepackage[acronym]{glossaries}
\usepackage{multirow}
\usepackage{tabularx, etoolbox, booktabs}
\usepackage{microtype}
\usepackage{pifont}
\usepackage{spconf}

\newacronym{BSS}{BSS}{blind source separation}
\newacronym{CNN}{CNN}{convolutional neural network}
\newacronym{ME}{ME}{mask estimator}
\newacronym{MSE}{MSE}{mean squared error}
\newacronym{DFT}{DFT}{discrete Fourier transform}
\newacronym{STFT}{STFT}{short-time Fourier transform}
\newacronym{ISTFT}{ISTFT}{inverse STFT}
\newacronym{SI-SDR}{si-SDR}{scale invariant signal-to-distortion ratio}
\newacronym{SDR}{SDR}{signal-to-distortion ratio}
\newacronym{TasNet}{TasNet}{time-domain audio separation network}
\newacronym{RSAN}{RSAN}{recurrent selective attention network}
\newacronym{ASR}{ASR}{automatic speech recognition}
\newacronym{MVDR}{MVDR}{minimum variance distortionless response}
\newacronym{MWF}{MWF}{multichannel wiener filter}
\newacronym{GEV}{GEV}{generalized eigenvalue}
\newacronym{WER}{WER}{word error rate}
\newacronym{BAN}{BAN}{blind analytic normalization}
\newacronym{NN}{NN}{neural network}
\newacronym{AM}{AM}{Acoustic Model}
\newacronym{PSD}{PSD}{power spectral density}
\newacronym{PDF}{PDF}{Probability Density Function}
\newacronym{ATF}{ATF}{acoustic transfer function}
\newacronym{RIR}{RIR}{room impulse response}
\newacronym{RNN}{RNN}{recurrent neural network}
\newacronym{SNR}{SNR}{Signal to Noise Ratio}
\newacronym{HMM}{HMM}{Hidden Markov Model}
\newacronym{GMM}{GMM}{Gaussian Mixture Model}
\newacronym{PSM}{PSM}{Phase Sensitive Mask}
\newacronym{tf}{tf}{time-frequency}
\newacronym{SB}{SB}{SpeakerBeam}
\newacronym{AUX}{AUX}{Auxiliary Network}
\newacronym{SCM}{SCM}{spatial correlation matrix}
\newacronym{RTF}{RTF}{relative transfer function}
\newacronym{MM}{MM}{mixture models}
\newacronym{DC}{DC}{deep clustering}
\newacronym{DAN}{DAN}{deep attractor network}
\newacronym{BiLSTM}{BiLSTM}{bidirectional long short term memory}
\newacronym{TCN}{TCN}{temporal convolutional network}
\newacronym{FF}{FF}{feed forward layer}
\newacronym{WSJ}{WSJ}{Wall Street Journal}
\newacronym{IBM}{IBM}{Ideal Binary Mask}
\newacronym{BCE}{BCE}{binary cross entropy}
\newacronym{PIT}{u-PIT}{utterance-level permutation invariant training}
\newacronym{OR-PIT}{OR-PIT}{one-and-rest permutation invariant training}

\usepackage[font={small}]{caption, subfig}
\title{DEMYSTIFYING TASNET: A DISSECTING APPROACH}
\name{Jens Heitkaemper, Darius Jakobeit, Christoph Boeddeker, Lukas Drude, Reinhold Haeb-Umbach}
\address{Paderborn University, 
		Department of Communications Engineering,
		Paderborn, Germany\\
		{\small \tt \{heitkaemper, boeddeker, drude, haeb\}@nt.uni-paderborn.de}}
\begin{document}
\ninept
\maketitle
\begin{abstract}
In recent years time domain speech separation has excelled over frequency domain separation in single channel scenarios and noise-free environments. 
In this paper we dissect the gains of the \gls{TasNet} approach by gradually replacing components of an \gls{PIT} based separation system in the frequency domain until the \gls{TasNet} system is reached, thus blending components of frequency domain approaches with those of time domain approaches.
Some of the intermediate variants achieve comparable \gls{SDR} gains to \gls{TasNet}, but retain the advantage of frequency domain processing: compatibility with classic signal processing tools such as frequency-domain beamforming and the human interpretability of the masks.
Furthermore, we show that the \gls{SI-SDR} criterion used as loss function in \gls{TasNet} is related to a logarithmic mean square error criterion and that it is this criterion which contributes most reliable to the performance advantage of \gls{TasNet}.
Finally, we critically assess which gains in a noise-free single channel environment generalize to more realistic reverberant conditions.

%
\end{abstract}
\section{Introduction}
\label{sec:intro}
In recent years multiple neural network based speech source separation techniques have been proposed. 
Techniques like  \gls{DC} \cite{Hershey16DC, Wang18Chimera}, \gls{DAN} \cite{Chen17DAN}, \gls{PIT} \cite{Yu17PIT, Kolbaek17UPit} and \gls{RSAN} \cite{Kinoshita18RSAN} have achieved remarkable single-channel separation performance on mixtures of unknown speakers.
All of the above mentioned separation systems rely on a transformation to the frequency domain, where separation masks are computed from the magnitude spectra. For the reconstruction of the separated time domain signals either the phase of the mixture signal is used, or dedicated phase reconstruction techniques are applied \cite{Roux19Phasebook, Wang2018EndtoEndSS}. The processing of complex-valued \gls{STFT} representations has also been considered \cite{Wang19PhaseReconstruction}. 

More recently, approaches have been introduced which transform the time domain signal to a non-negative real-valued, learned latent space to do source separation. These techniques naturally account for the phase, thus rendering explicit phase reconstruction superfluous. Examples for those approaches are \gls{TasNet} \cite{Luo2018TasNetSI}, Conv-\gls{TasNet} \cite{Yi19ConvTasNet}, FurcaNet \cite{Shi19FurcaNet} and \gls{OR-PIT} \cite{Takahashi19OrPit}.
We stick here to the common terminology of naming them ``time domain'' techniques, although the separation actually takes place in a latent space to which the time domain signal is transformed.
Those approaches have been shown to achieve superior results in terms of separation performance compared to the frequency domain separation systems. However, results were only reported for single-channel data, and under noise-free and non-reverberant conditions.

In \cite{Bahmaninezhad2019TasNetSpect} a comparison of time and frequency domain separation has been done. The authors
found the \gls{SI-SDR} loss computed in the time domain to be a superior criterion compared to the frequency domain \gls{MSE} typically used in frequency domain source separation. But these observations could not fully explain the performance advantage of the time domain approach over frequency domain separation.

In this paper we start with the common \gls{PIT} frequency domain separation. We then gradually make the following modifications towards the time domain convolutional \gls{TasNet} structure:
\begin{compactitem}
  \item The magnitude spectrum representation is replaced by the real and imaginary part of the \gls{STFT}, thus capturing phase information
  \item The \gls{MSE} loss in the frequency domain is replaced with a \gls{SI-SDR} loss in the time domain
  \item Either the \gls{STFT} or the \gls{ISTFT} or both are replaced by learned transformations
	\item The frame resolution is reduced from \SI{64}{ms} frame width and \SI{16}{ms} frame advance to \SI{4}{ms} and \SI{2}{ms} as is used in TasNet
\end{compactitem}
The latter turns out to be a major source of performance improvement.
Additionally, we propose a reformulation of the \gls{SI-SDR} criterion which reveals that it is closely related to a logarithmic MSE loss.

Some of those ``hybrid'' time and frequency domain techniques achieve remarkable performance. For example, comparable separation performance to \gls{TasNet} can be achieved if either the encoder or the decoder are fixed to be the STFT or ISTFT. This opens the way to employ common frequency domain beamforming techniques for source extraction, which have been shown to deliver superior results to mask-based source extraction in many studies \cite{Yin2018MultitalkerSS,Drude2018SpecialIssue,Zmolikova17SpeakerBeam,Higuchi16RobustMVDR}.
Finally, the fixed STFT-based encoder allows for a human interpretability of the masks, while at the same time maintaining a separation close to the excellent performance of \gls{TasNet}. 

Another objective of this paper is to study the behavior of time domain separation techniques in (somewhat) noisy and reverberant environments.
There have been some initial examinations of this issue for example in \cite{Bahmaninezhad2019TasNetSpect}, \cite{Luo2019FaSNet} and \cite{Wichern2019WHAM}.
However, either they do not consider reverberant conditions \cite{Wichern2019WHAM}, do not achieve competitive results \cite{Luo2019FaSNet}, or they use oracle information to guide their systems as in \cite{Bahmaninezhad2019TasNetSpect}.
In \cite{Gu19TasNetSpatialFeatures} learnable spatial features for time domain separation are introduced and are shown to achieve superior \gls{SDR} results compared to frequency based separation systems.
In contrast, we evaluate the transferability of gains on noise-free single channel to reverberated data without changes in the network structure.




The remainder of the paper is structured as follows. In \cref{sec:signal_model} a generic view on both frequency and time domain source separation is introduced. Its elements are discussed in the following subsections. In \cref{sec:evaluation} the evaluation results are presented, and the paper concludes with a short discussion in \cref{sec:conclusion}.

\section{PIT Based Source Separation}
\label{sec:signal_model}
We consider a microphone signal $y(t)$ in the time domain, which can be represented as the sum of $K$ speech signals $x_k(t)$, and additive noise $n(t)$
\begin{align}
y(t) = \sum_{k=1}^K x_{k}(t) + n(t),
\end{align}
where $t$ is the time index.

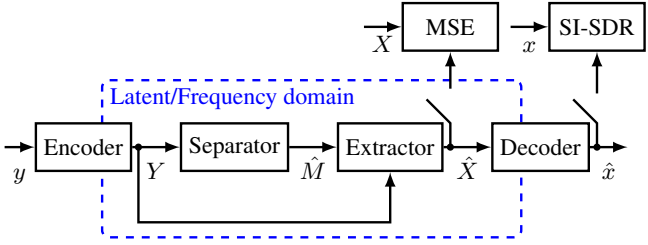
\begin{figure}[t]
	\tikzset{%
  block/.style    = {draw, line width=0.1em, rectangle, minimum height = 2em, minimum width = 2.5em, fill=white},
  sum/.style      = {draw, circle, node distance = 2cm}, 
  cross/.style={path picture={\draw[black](path picture bounding box.south east) -- (path picture bounding box.north west)
		 (path picture bounding box.south west) -- (path picture bounding box.north east);}},
               }
\tikzstyle{branch}=[{circle,inner sep=0pt,minimum size=0.3em,fill=black}]
\tikzstyle{box} = [draw, dashed, inner xsep=4mm, inner ysep=4mm, line width=0.1em, rounded corners=0.3em]
\tikzstyle{every path}=[line width=0.1em]
\tikzstyle{arrow} = [-latex]

\begin{tikzpicture}[auto, line width=1em, node distance = 1cm]
\pgfdeclarelayer{background}
\pgfdeclarelayer{foreground}
\pgfsetlayers{background,main,foreground}
\begin{pgfonlayer}{foreground}
	\node[block, align=center](encoder){Encoder};
	\node[block, align=center, right=2em of encoder](separator){Separator};
	\node[block, right = 2em of separator](mult){Extractor};
	\node[block, align=center, right=2em of mult](decoder){Decoder};
	
	\coordinate(pit) at ($ (mult) !0.5! (decoder) + (-0.2,1.6) $);
	\coordinate(sisnr_x) at ($ (decoder) + (0.75,0) $);
	\coordinate(sisnr_cord) at ($ ( sisnr_x |- pit) $);
	\node[block, align=center, minimum width = 4em, at=(pit)](mse){MSE};
	\node[block, align=center, minimum width = 4em, at=(sisnr_cord)](sisnr){SI-SDR};

	\draw[arrow] ($(encoder.west)  + (-0.4, 0)$) -- node[below]{$\vphantom{\hat{X}}y$} (encoder.west);
	\draw[arrow] (encoder.east) -- node[below]{$\vphantom{\hat{X}}Y$} (separator.west);
	\draw[arrow] (separator.east) -- node[below]{$\hat{M}$} (mult.west);
	\draw[arrow] (mult.east) -- node[below]{$\hat{X}$} (decoder.west);
	\draw[-] (mult.east -| mse.south) node[branch]{} -- ($ (mse.south) + (0,-3em) $) -- ($ (mse.south) + (-1em,-2em) $);
	\draw[arrow] ($ (mse.south) + (0,-1.55em) $) -- (mse.south);
	\draw[arrow] (decoder.east) -- node[below]{$\vphantom{\hat{X}}\hat{x}$} ($(decoder.east) + (0.5,0) $);
	\draw[-] (decoder.east -| sisnr.south) node[branch]{} -- ($ (sisnr.south) + (0,-3em) $) -- ($ (sisnr.south) + (-1em,-2em) $);
	\draw[arrow] ($ (sisnr.south) + (0,-1.75em) $) -- (sisnr.south);
	\draw[arrow] ($(mse.west)  + (-0.5, 0)$) -- node[below]{$X$} (mse.west);
	\draw[arrow] ($(sisnr.west)  + (-0.5, 0)$) -- node[below]{$\vphantom{X}x$} (sisnr.west);
	\draw[arrow] ($ (separator.west) + (-0.55,0) $) node[branch]{} |- ($ (separator.west) + (0,-1) $) -| (mult.south);
\end{pgfonlayer}
\begin{pgfonlayer}{background}
	\coordinate(tmp) at ($(separator) + (0, -0.8)$);
	\coordinate(tmp2) at ($(separator.north) + (0, 0.5em)$);
	\node (box) [box, blue, text=blue, fit=(encoder.east) (separator) (tmp) (tmp2) (decoder.west), label={[anchor=north west]north west:\color{blue}Latent/Frequency domain}] {};
\end{pgfonlayer}
\end{tikzpicture}
	\caption{Block diagram of the PIT-based source separation.}
	\label{fig:block}
\end{figure}

\Cref{fig:block} offers a generic view on a source separation system. It consists of an encoder which transforms the input signal into a domain suitable for source separation, a mask estimation separator network which estimates a mask for each source, a source extractor, which outputs an estimate of each source signal in the transform domain, and a decoder which transforms  the extracted source signals  back to the time domain.
Encoder and decoder can be either learned transforms, as in time domain separation, or they can be fixed to be STFT and ISTFT as in frequency domain processing. The mask estimator can be realized by a network of BLSTM or CNN layers, and source extraction can be achieved either by mask multiplication or, in case of multi-channel frequency domain processing, by beamforming.

\glsreset{PIT}
\noindent Dependent on the system the loss will be either calculated in the frequency or in the time domain.
In both scenarios the loss is calculated for each permutation and the best permutation is used which is called \gls{PIT} \cite{Kolbaek17UPit}.

\subsection{Encoder/Decoder}
\label{sec:encoder_decoder}

As part of the transformation the encoder segments the input signal $y(t)$ into $L$ segments (frames) of length $L_\mathrm{w}$, with an overlap of $L_\mathrm{w}-L_\mathrm{s}$, where $L_\mathrm{s}$ is the advance  between the segments. The segmented observation is represented by $y_i(\ell) := y(i+\ell\cdot L_\mathrm{s})$, where $\ell$ counts the segments and $i$ the samples in a segment.

The most common encoder for source separation with \gls{PIT} is the \gls{STFT} \cite{Kolbaek17UPit, Yin2018MultitalkerSS}
\begin{align}
	Y(\ell, f) &= \sum_{i=0}^{L_\mathrm{w}} y_i(\ell)\cdot w_i\cdot \e^{-\im 2\pi\cdot i \cdot f /F_\text{DFT} }.
\end{align}
Here, $w_i$ is a fixed window function and $F_\text{DFT}$ is the size of the \gls{DFT}, which need not be identical to $L_\mathrm{w}$.
The feature index $f$ takes values in the range from $0$ to $F-1$.
It is common to ignore redundant frequency bins and set the number of features to $F=F_\text{DFT} / 2 + 1$.
The corresponding decoder is the \gls{ISTFT}.

Alternatively, a one-dimensional \gls{CNN} is used for both the encoder and decoder to transform the signal to a latent space
\begin{align}
	Y(\ell, f) &= \mathrm{ReLU}\left( \sum_i y_i(\ell) u_i(f)\right),
\end{align}
and back with
\begin{align}
	\hat{x}_{k}(t) &= \sum_\ell \mathrm{rect}\left(\frac{t- \ell L_\mathrm{s}-\frac{L_\mathrm{w}}{2}}{L_\mathrm{w}}\right) \sum_{f} \tilde{u}_{i=t-\ell L_\mathrm{s}}(f) \hat{X}_k(\ell,f)
\end{align}
where $\hat{x}_{k}(t)$ is the time domain estimate of the signal of speaker $k$, while $u_i(f)$ and $\tilde{u}_i(f)$ are learnable one-dimensional kernels of the encoder and decoder, respectively. $\mathrm{ReLU}$ symbolizes the rectified linear unit non-linearity.

Note that the (I)STFT can also be written as one-dimensional convolution operations, however with  fixed instead of learned kernel functions. 


Independent of which of the above encoders is used, the separator module assumes that the signals superpose linearly in the latent domain:
\begin{align}
Y(\ell,f) = \sum_{k} X_{k}(\ell,f) + N(\ell,f)
\end{align}
where $X_{k}$ and $N$ are defined similarly as $Y$.
Although this assumption is only true, when the encoder is linear, the entire training procedure encourages the encoder to find a representation in which linear demixing leads to a good separation performance.

\subsection{Mask Estimation Separator Network}
\label{sec:separator}
In case of the \gls{STFT} as encoder it is common to compute the magnitude spectrum and present it to the separation network. To maintain phase information, one can alternatively present the concatenation of the real and imaginary part of the STFT at the mask estimator input, and the  network is trained to compute two masks per time frequency bin, one for the real and one for the imaginary part of the frequency domain representation.

The mask estimation layers of many frequency domain separation systems consist of recurrent layers \cite{Hershey16DC, Kolbaek17UPit, Chen17DAN}.
However, in the time domain approach of \cite{Yi19ConvTasNet} convolution layers have been shown to lead to superior separation results.
This work will focus on convolution layers and, as a side effect, show that they can also be used to improve separation results in the frequency domain.

\subsection{Cost function}
\label{sec:loss}
The loss for network training can either be calculated in the time or in the transform domain.

In the frequency domain, using magnitude spectra at the input of the mask estimation network, the  phase sensitive \gls{MSE} is the most common loss function for permutation invariant training \cite{Kolbaek17UPit,Wichern2019WHAM}:
\begin{align}
\mathcal{L}^{\mathrm{PMSE}} = \frac{1}{L\cdot F\cdot K}\sum_{\ell,f,k}\abs{\abs{\hat{X}_k} - \abs{X_k}\cdot\cos(\Delta \theta_k)}^2
\end{align}
where $\Delta \theta_k = \theta_y-\theta_k$ is the difference between the mixture phase $\theta_y$  and the phase $\theta_k$ of the clean speech signal of speaker $k$.
If the separator estimates two separate masks for the real and imaginary part, the simple \gls{MSE} is used
\begin{align}
\mathcal{L}^{\mathrm{MSE}} = \frac{1}{L\cdot F\cdot K}\sum_{\ell,f,k}|\hat{X}_k - X_k|^2.
\end{align}
In time domain source separation, the \gls{SI-SDR} is rather common~\cite{Luo2018TasNetSI}: 
\begin{align}
 \label{eq:si-sdr}
\mathcal{L}^{\mathrm{SI-SDR}} &= -10\frac{1}{K} \sum_{k} \log_{10}\frac{\sum_t\abs{\alpha \cdot x(t)}^2}{\sum_t\abs{\alpha\cdot x(t)-\hat{x}(t)}^2}\\
\text{with } \alpha &= \frac{\sum_t \hat{x}(t)\cdot x(t)}{\sum_t\abs{x(t)}^2}.
\end{align}
Using $\beta$ = $\frac{1}{\alpha}$ this loss function can be written as
\begin{align}
\mathcal{L}^{\mathrm{SI-SDR}} &= -10\frac{1}{K} \sum_{k} \log_{10}\frac{\sum_t\abs{x(t)}^2}{\sum_t\abs{x(t)-\beta \cdot \hat{x}(t)}^2} \nonumber \\
  \label{eq:log-mse-1}
 &\propto 10\frac{1}{K} \sum_{k} \log_{10}\sum_t\abs{x(t)-\beta \cdot \hat{x}(t)}^2,
\end{align}
where we removed all terms not dependent on learnable parameters in \eqref{eq:log-mse-1}.
Setting the weighting factor $\beta=1$ results in  a logarithmic \gls{MSE} loss
\begin{align}
  \label{eq:log-mse}
	\mathcal{L}^{\mathrm{T-LMSE}}&=10\frac{1}{K} \sum_{k} \log_{10}\sum_t\abs{x(t)-\hat{x}(t)}^2.
\end{align}
The \gls{MSE} can further be motivated by the logarithmic perception of loudness of the human hearing: less loud regions can still influence perception quite substantially.

The main difference between the two cost function in Equation \eqref{eq:si-sdr} and Equation \eqref{eq:log-mse} is the scaling of the estimate, which appears to be a minor difference. Therefore, both loss functions are expected to achieve similar source separation results.
\gls{PIT} is used to solve the global permutation problem for all cost functions, be it computed in the frequency or in the time domain.

\subsection{Source Extraction}
\label{sec:beamforming}
With single-channel input, the actual source extraction is achieved by multiplying the transformed microphone signal with the speaker specific masks obtained from the mask estimation network: 
\begin{align}
\hat{X}_k(\ell,f)=Y(\ell,f)\cdot \hat{M}_k(\ell,f),\quad k \in \{1,\ldots, K\}.
\end{align}
%
If, however, multi-channel data is available, beamforming is known to extract source signals with less artifacts than masking.
Although mask-based beamforming in the frequency domain is used in many current neural network based systems \cite{Yin2018MultitalkerSS,Drude2018SpecialIssue,Zmolikova17SpeakerBeam,Higuchi16RobustMVDR} we have chosen to use time-domain beamforming since it is more compatible with the \gls{TasNet} structure:
\begin{align}
\hat{x}^\text{BF}_k(t) = \mathbf{f}_k(t)^\mathsf{T} \tilde{\mathbf{y}}(t),
\end{align}
with $\mathbf{y}(t) = [y_1(t-L_\mathrm{f}), ..., y_D(t-L_\mathrm{f}), ..., y_1(t), ..., y_D(t)]^\mathsf{T}$ being the multi-channel microphone signal stacked with $L_\mathrm{f}$ values of the history, where $D$ denotes the number of channels.

For the computation of the vector of beamformer coefficients $\mathbf{f}_k$, several criteria exist.
Here $\mathbf{f}_k$ is estimated using a \gls{MSE} based criterion similar to \cite{Zhang2017GlottalMB}
\begin{align}
\mathbf{f}_k = \underset{\mathbf{f}_k}{\text{argmin}}\ExpOp{\left[\abs{x_k(t) -\mathbf{f}_k(t)^\mathsf{T}\mathbf{y}(t)}^2\right]},
\end{align}
which is a Wiener filtering problem with the solution
\begin{align}
\mathbf{f}_k = \ExpOp\left[\mathbf{y}(t)\mathbf{y}(t)^\mathsf{T}\right]^{-1}\ExpOp\left[\mathbf{y}(t)x_k(t)\right],
\end{align}
where $x_k(t)$ is approximated with the decoder output $\hat{x}_k(t)$ at a reference channel and the expectation with an average over the time.



\section{Evaluation}
\label{sec:evaluation}
The separation systems are compared on two databases using the BSSEval \gls{SDR} \cite{Raffel14Mireval}, \gls{SI-SDR} and \gls{WER} as metrics.
For \gls{WER} calculation we use a baseline Kaldi model \cite{Povey11Kaldi} trained on the \gls{WSJ} database \cite{Garofolo1993WSJ} without retraining and language model rescoring.
All experiments use the convolutional separator architecture described in \cite{Yi19ConvTasNet}  with a ReLU activation and global layer normalization. 
The hyperparameter $F=514$ and $F_{\mathrm{DFT}}=512$ are constant for all experiments if not otherwise specified.
We use Adam with a step size of $\alpha=0.001$ which is divided by $2$ if the validation metric does not improve for $10$ epochs.

\subsection{Database}
\label{sec:database}
Two databases are used for evaluation.
The first is the WSJ0-2mix database \cite{Hershey16DC} where two utterances from the \gls{WSJ} database spoken by different speakers are mixed at random \gls{SDR} between \SI{-2.5}{dB} and \SI{2.5}{dB}.
The database is split into \SI{30}{h} of training data, \SI{10}{h} of validation data and \SI{5}{h} of testing data.
As length of each example the length of the longest utterance was chosen to allow \gls{WER} estimation for both speaker.
Most published results choose the length of the example to be equal to the length of the shorter utterance which leads to about \SI{0.5}{dB} higher \gls{SDR} results compared to our definition.

The second database is a newly released spatialized multi-speaker database \cite{Drude18SMSWSJ}, which consists of utterances from the \gls{WSJ} database convolved with  simulated \glspl{RIR} mixed with an utterance of a different speaker convolved with a different \gls{RIR} at a random \gls{SDR}.
Additionally, white noise with a \gls{SNR} between $20$ and \SI{30}{dB} is added to the mixture.

\subsection{From Frequency to Time Domain Source Separation}
\label{sec:separation}
In \cref{tbl:freq_time} the different steps from a common frequency domain \gls{PIT} model to a Conv-\gls{TasNet} are shown on the single channel database.
The domain names "Frequ-Abs" and "Frequency" represent the frequency magnitude spectrum and the frequency spectrum split into its real and imaginary part, respectively.

 \begin{table}[b]
 	\vspace*{-0.35cm}
 	\centering
 	\caption{A step-by-step comparison of frequency and time domain source separation on the test set of the WSJ0-2mix database.}
 	\label{tbl:freq_time}
 	\begin{tabular}{
 			l
 			c
 			l
 			S[table-auto-round, table-format=2.1]
 			S[table-auto-round, table-format=2.1]
 			S[table-auto-round, table-format=2.2]} \toprule
 		Domain & $L_\mathrm{w}$/$L_\mathrm{s}$ & Loss-Fn &{\gls{SI-SDR}} & {\gls{SDR}} & {\gls{WER}}\\
 		& ms/ms & &  {dB} & {dB} & {$\%$}\\ \toprule
 		Frequ-Abs & 64/16 & $\mathcal{L}^\mathrm{PMSE}$ & 8.768591 & 9.275271 &39.45\\
 		Frequ-Abs & 64/16 & $\mathcal{L}^\mathrm{MSE}$ & 7.381150 & 7.815363 & 41.03 \\
 		Frequency & 64/16 & $\mathcal{L}^\mathrm{MSE}$ & 2.310123 & 4.612538 & 51.71\\
 		Frequency & 64/16 & $\mathcal{L}^\mathrm{SI-SDR}$ & 5.753887 & 6.238750 & 50.72 \\
 		Frequency & 4/2 & $\mathcal{L}^\mathrm{SI-SDR}$ & 12.431684 & 12.778539 & 24.69 \\
 		Latent & 4/2 & $\mathcal{L}^\mathrm{SI-SDR}$ & 14.361643 & 14.736700 & 21.71\\
 		\bottomrule
 	\end{tabular}
 \end{table}

In \cite{Kolbaek17UPit} a BLSTM based separation system using the \gls{PIT} loss achieves a \gls{SDR} of \SI{9.3}{dB}.
Using the \gls{TCN} separator we get similar results so that all experiments will be done using the \gls{TCN}.
Removing the phase information from the loss reduces the \gls{SDR} to \SI{7.8}{dB} and after splitting the frequency spectrum in the real and imaginary part the results decrease further to \SI{4.6}{dB}.
Interestingly, the change from the frequency to the time domain loss only results in a small gain.
Whereas, the switch to the shorter $L_\mathrm{w}$ and $L_\mathrm{s}$ improve the results by a large margin of \SI{6.6}{dB}.
The last change required at a Conv-\gls{TasNet} model is to do the separation in the latent domain instead of in the frequency domain which leads to a further improvement of \SI{1.9}{dB} in \gls{SDR}.

\begin{table}[t]
	\vspace*{-0.35cm}
	\centering
	\caption{Comparison of different time domain loss functions using $L_\mathrm{w}=\SI{4}{ms}$ and $L_\mathrm{s}=\SI{2}{ms}$ with for trainable encoder/decoder on the test set of the WSJ0-2mix database.}
	\label{tbl:loss}
	\begin{tabular}{
			l
			S[table-auto-round, table-format=2.1]
			S[table-auto-round, table-format=2.1]
			S[table-auto-round, table-format=2.2]} \toprule
		Loss-Fn & {\gls{SI-SDR}} & {\gls{SDR}} & {\gls{WER}}\\
		& {dB} & {dB} & {$\%$}\\ \toprule
		$\mathcal{L}^\mathrm{SI-SDR}$&14.361643 & 14.736700 & 21.71 \\
		$\mathcal{L}^\mathrm{T-LMSE}$ & 14.43421 & 14.926137 & 21.46 \\
		$\mathcal{L}^\mathrm{T-MSE}$ & 10.901169 & 11.310080 & 29.43 \\
		\bottomrule
	\end{tabular}
\end{table}
In \cref{tbl:loss} the presented time domain loss functions are compared for trainable encoder/decoder.
The T-LMSE loss achieves similar results to the \gls{SI-SDR} loss as expected.
Removing the logarithmic scaling in T-MSE reduces the \gls{SDR} by more than \SI{2}{dB}.

In this section we have shown that the main reasons for the superior results of the \gls{TasNet} is not the use of a learned latent space but the time domain loss function and the small window and shift size.
Therefore, both will be used in the following experiments.
 
\subsection{Encoder/Decoder Comparison}
In \cref{tbl:encoder_decoder} the different encoder and decoder combinations introduced in \cref{sec:signal_model} are compared on the WSJ0-2mix database.
Here all experiments use the smaller window and shift size since they achieved the best results in the previous section \ref{sec:separation}.

 \begin{table}[b]
 	\vspace*{-0.35cm}
 	\centering
 	\caption{Comparison of different encoder and decoder combination using $L_\mathrm{w}=\SI{4}{ms}$ and $L_\mathrm{s}=\SI{2}{ms}$ on the test set of the WSJ0-2mix database.}
 	\label{tbl:encoder_decoder}
 	\begin{tabular}{
 			l
 			c
 			c
 			S[table-auto-round, table-format=2.1]
 			S[table-auto-round, table-format=2.1]
 			S[table-auto-round, table-format=2.2]} \toprule
 		Loss-Fn & Encoder & Decoder & {\gls{SI-SDR}} & {\gls{SDR}} & {\gls{WER}}\\
 		& & & {dB} & {dB} & {$\%$}\\ \toprule
 		$\mathcal{L}^\mathrm{SI-SDR}$ & learned & learned & 14.361643 & 14.736700 & 21.71 \\
 		$\mathcal{L}^\mathrm{SI-SDR}$ & \gls{STFT} & learned & 13.934528 & 14.301157 & 21.92 \\
 		$\mathcal{L}^\mathrm{SI-SDR}$ & learned & \gls{ISTFT} & 14.119177 & 14.507232 & 21.87 \\
 		$\mathcal{L}^\mathrm{SI-SDR}$ & \gls{STFT} & \gls{ISTFT} & 12.431684 & 12.778539 & 24.69\\
 		\bottomrule
 	\end{tabular}
 \end{table}
 
The classical learned encoder/decoder structures as used in the original \gls{TasNet} achieves a \gls{SDR} of \SI{14.7}{dB} which is similar to the number presented in \cite{Yi19ConvTasNet}.
Replacing the encoder or the decoder with their \gls{STFT} counterpart leads only to a slight reduction of the separation results.
This is especially interesting for the combination of learned encoder and \gls{ISTFT} since the encoder includes the RelU non-linearity and therefore just outputs positive values which is not matching with the \gls{STFT} output.
One explanation for these results are the redundancies introduced by the gap between the window length $L_\mathrm{w}$ and the size of the feature vector $F$ which offer the separation a high degree of freedom.
When replacing both the encoder and decoder with their \gls{STFT} counterpart the \gls{SDR} is reduced by \SI{2}{dB}.


\begin{figure}[t]
	\centering
	\newlength\figureheight
	\newlength\figurewidth
	\setlength\figureheight{4.8cm}
	\setlength\figurewidth{3.3cm}
\begin{tikzpicture}
\pgfplotsset{every axis/.append style={
		xlabel style={font=\footnotesize},
		ylabel near ticks,
		ylabel style={font=\footnotesize},
		tick label style={font=\footnotesize},
		title style={font=\footnotesize},
		colorbar style={ylabel={Energy / dB}, yticklabel pos=right, width=0.05\figurewidth},
	}}
\begin{axis}[
name=plot1,
point meta max=15.6704769134521,
point meta min=-44.3295249938965,
tick align=outside,
tick pos=left,
title={\!\!\!\!STFT/ISTFT\!\!\!\!},
width=\figurewidth,
height=\figureheight,
x grid style={white!69.01960784313725!black},
xlabel={Segment $\ell$},
xtick={0, 500},
xmin=-0.5, xmax=700.5,
xtick style={color=black},
y grid style={white!69.01960784313725!black},
ylabel={Latent vector index},
ymin=-0.5, ymax=513.5,
ytick style={color=black},
ytick={0, 256, 512},
yticklabels ={0, 256, 512},
]
\addplot graphics [
xmin=-0.5, xmax=700.5,
ymin=-0.5, ymax=513.5,
includegraphics={trim=200 0 1344 0, clip}  
] {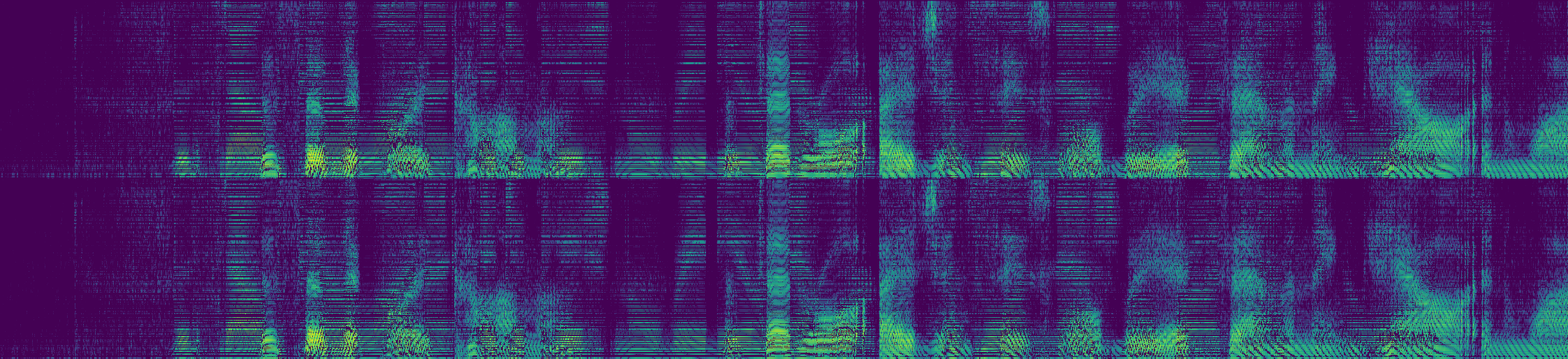};
\end{axis}
\begin{axis}[
name=plot2,
colormap/viridis,
at=(plot1.right of south east), anchor=south west,
xshift=1mm,
point meta max=15.6704769134521,
point meta min=-44.3295249938965,
tick align=outside,
tick pos=left,
title={\!\!\!\!STFT/learned\!\!\!\!},
height=\figureheight,
width=\figurewidth,
x grid style={white!69.01960784313725!black},
xlabel={Segment $\ell$},
xtick={0, 500},
xmin=-0.5, xmax=700.5,
xtick style={color=black},
y grid style={white!69.01960784313725!black},
ytick={0, 256, 512},
yticklabels ={},
ymin=-0.5, ymax=513.5,
ytick style={color=black},
]
\addplot graphics [
xmin=-0.5, xmax=700.5,
ymin=-0.5, ymax=513.5,
includegraphics={trim=200 0 1344 0, clip}  
] {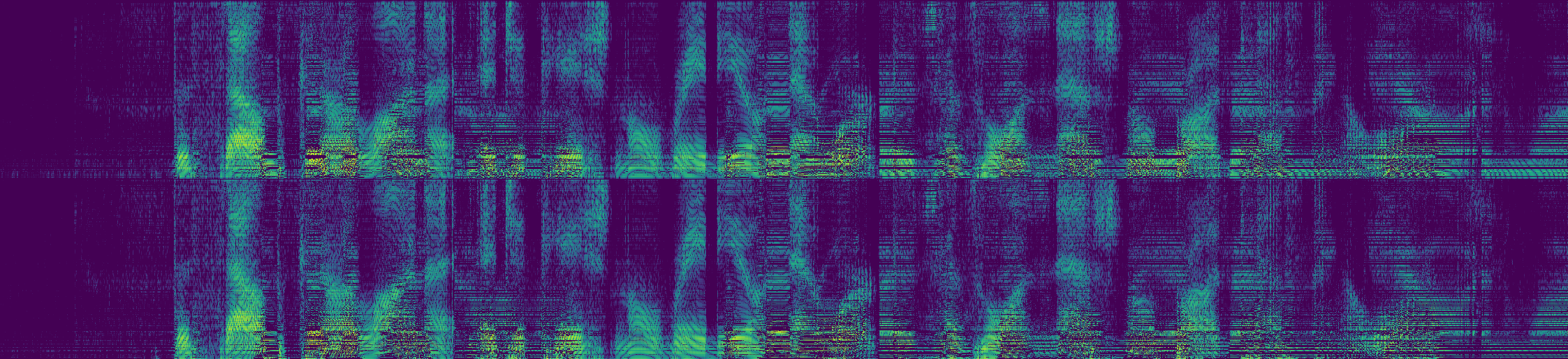};
\end{axis}
\begin{axis}[
name=plot3,
at=(plot2.right of south east), anchor=south west,
xshift=1mm,
point meta max=13.2472333908081,
point meta min=-46.7527694702148,
tick align=outside,
tick pos=left,
title={\!\!\!\!learned/ISTFT\!\!\!\!},
height=\figureheight,
width=\figurewidth,
x grid style={white!69.01960784313725!black},
xlabel={Segment $\ell$},
xtick={0, 500},
xmin=-0.5, xmax=700.5,
xtick style={color=black},
y grid style={white!69.01960784313725!black},
ymin=-0.5, ymax=513.5,
ytick style={color=black},
ytick={0, 256, 512},
yticklabels ={},
]
\addplot graphics [
xmin=-0.5, xmax=700.5,
ymin=-0.5, ymax=513.5,
includegraphics={trim=200 0 1344 0, clip}  
] {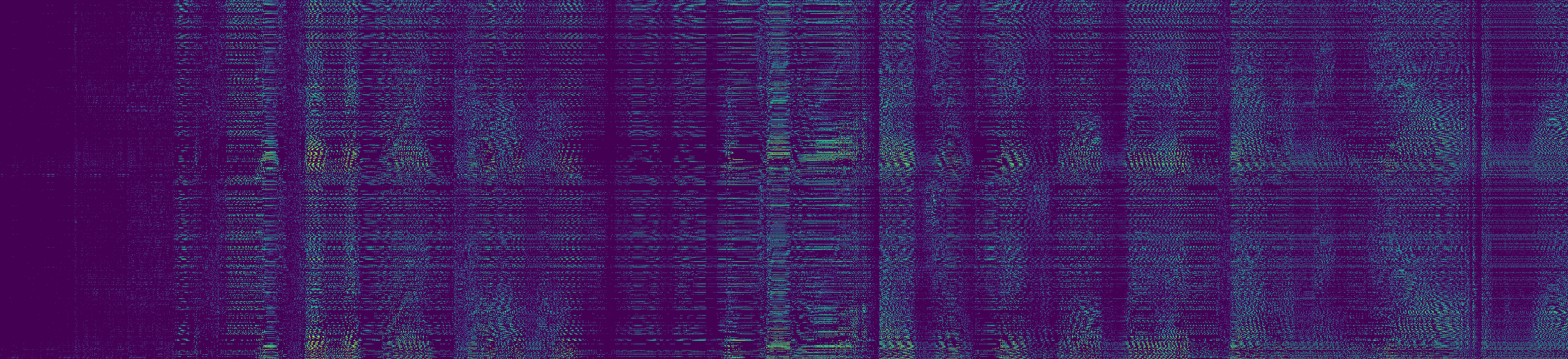};
\end{axis}
\begin{axis}[
at=(plot3.right of south east), anchor=south west,
xshift=1mm,
colormap/viridis,
point meta max=9.46504974365234,
point meta min=-50.5349502563477,
tick align=outside,
tick pos=left,
title={\!\!\!\!learned/learned\!\!\!\!},
height=\figureheight,
width=\figurewidth,
x grid style={white!69.01960784313725!black},
xlabel={Segment $\ell$},
xtick={0, 500},
xmin=-0.5, xmax=700.5,
xtick style={color=black},
y grid style={white!69.01960784313725!black},
ytick={0, 256, 512},
yticklabels ={},
ymin=-0.5, ymax=513.5,
ytick style={color=black}
]
\addplot graphics [
xmin=-0.5, xmax=700.5,
ymin=-0.5, ymax=513.5,
includegraphics={trim=200 0 1344 0, clip}  
] {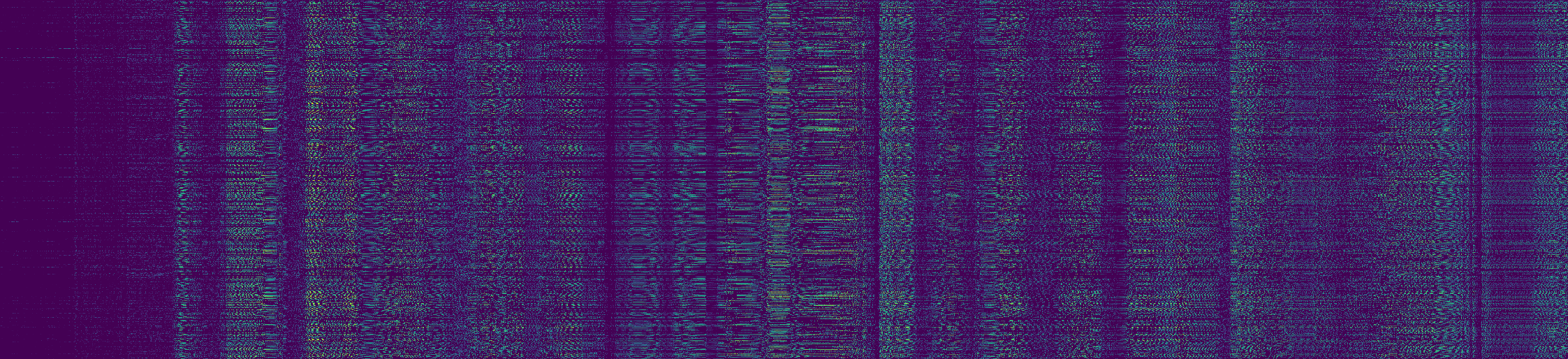};
\end{axis}

\end{tikzpicture}
    \vspace{-4mm}
	\caption{Example enhanced signals in the latent space using different encoder, decoder combinations. Color encodes energy in each bin from \SI{-50}{dB} (blue) to \SI{10}{dB} (yellow).}
	\label{fig:masks}
	\vspace{-1.5em}
\end{figure}

However, this reduction in \gls{SDR} may be acceptable given the human interpretability of the estimated masks and the opportunity to use further statistical enhancement algorithms.
\cref{fig:masks} shows the enhanced signal before decoding by the different encoder-decoder combinations.
Clearly, when using the STFT as encoder the spectrograms look as expected.
While for a learned encoder with ISTFT decoder one can see some familiar structures, the results for the case of learned encoder and decoder are very difficult to interpret.
\vspace{-0.5em}
\subsection{Time Domain Separation in Reverberant Environment}
In \cref{tbl:freq_time_rev} the same steps from a frequency domain \gls{PIT} system to the Conv-\gls{TasNet} as in \cref{tbl:freq_time} are evaluated on the simulated reverberant database described in \cref{sec:database}.

All models are retrained on the reverberant database and the results are achieved using time domain beamforming as described in \cref{sec:beamforming}.
 
  \begin{table}[b]
	\centering
	\caption{A step-by-step comparison of frequency and time domain source separation on the test set of a simulated reverberant database.}
	\label{tbl:freq_time_rev}
	\begin{tabular}{
			l
			c
			l
			S[table-auto-round, table-format=1.2]
			S[table-auto-round, table-format=2.2]} \toprule
		Domain & $L_\mathrm{w}$/$L_\mathrm{s}$ & Loss-Fn & {\gls{SDR}} & {\gls{WER}}\\
		& ms/ms & &  {dB} & {$\%$}\\ \toprule
		Frequ-Abs & 64/16 & $\mathcal{L}^\mathrm{PMSE}$ & 8.634910 & 38.25\\
		Frequ-Abs & 64/16 & $\mathcal{L}^\mathrm{MSE}$ & 8.364753  & 39.60 \\
		Frequency & 64/16 & $\mathcal{L}^\mathrm{MSE}$ & 6.046798 & 58.58 \\
		Frequency & 64/16 & $\mathcal{L}^\mathrm{SI-SDR}$ & 9.127018 & 35.36  \\
		Frequency & 4/2 & $\mathcal{L}^\mathrm{SI-SDR}$ & 6.721578 & 48.18 \\
		Latent & 4/2 & $\mathcal{L}^\mathrm{SI-SDR}$ & 6.505271 & 47.65\\
		Latent & 64/16 & $\mathcal{L}^\mathrm{SI-SDR}$ & 7.808781 & 43.39 \\
		\bottomrule
	\end{tabular}
\end{table}

The results show some remarkable differences to the single-channel results of the earlier sections.
Neither the higher frequency resolution nor the use of a latent space lead to an improvement. 
The evaluation implies that the training of a latent space for the separation of reverberated data is not trivial and that it maybe worthwhile to further investigate frequency domain separation.
 

\section{Conclusion}
\label{sec:conclusion}
In this paper a detailed dissection of \gls{TasNet} and a thorough investigation of its gains are presented.
We show empirically that on a noise-free single channel database the gains can be attributed to a high time resolution and the time domain loss.
However, in a reverberant environment only the time domain loss still leads to an improved separation.
Additionally, we introduce a logarithmic \gls{MSE} loss which improves upon the results achieved with the \gls{SI-SDR} loss.
For future work we aim to achieve a stronger combination of the time domain loss and frequency domain separation which may allow the use of well known frequency domain extraction methods like statistical beamforming.
\section{Acknowledgment}
This work was supported by DFG under contract number Ha3455/14-1.
Computational resources were provided by the Paderborn Center for Parallel Computing.
\clearpage
\balance
\bibliographystyle{IEEEbib}
\bibliography{mybib}

\begin{thebibliography}{10}

\bibitem{Hershey16DC}
J.~R. {Hershey}, Z.~{Chen}, J.~{Le Roux}, and S.~{Watanabe},
\newblock ``Deep clustering: Discriminative embeddings for segmentation and
  separation,''
\newblock in {\em 2016 IEEE International Conference on Acoustics, Speech and
  Signal Processing (ICASSP)}, March 2016, pp. 31--35.

\bibitem{Wang18Chimera}
Z.~{Wang}, J.~{Le Roux}, and J.~R. {Hershey},
\newblock ``Alternative objective functions for deep clustering,''
\newblock in {\em 2018 IEEE International Conference on Acoustics, Speech and
  Signal Processing (ICASSP)}, April 2018, pp. 686--690.

\bibitem{Chen17DAN}
Z.~{Chen}, Y.~{Luo}, and N.~{Mesgarani},
\newblock ``Deep attractor network for single-microphone speaker separation,''
\newblock in {\em 2017 IEEE International Conference on Acoustics, Speech and
  Signal Processing (ICASSP)}, March 2017, pp. 246--250.

\bibitem{Yu17PIT}
D.~{Yu}, M.~{Kolbaek}, Z.~{Tan}, and J.~{Jensen},
\newblock ``Permutation invariant training of deep models for
  speaker-independent multi-talker speech separation,''
\newblock in {\em 2017 IEEE International Conference on Acoustics, Speech and
  Signal Processing (ICASSP)}, March 2017, pp. 241--245.

\bibitem{Kolbaek17UPit}
M.~{Kolbaek}, D.~{Yu}, Z.~{Tan}, and J.~{Jensen},
\newblock ``Multitalker speech separation with utterance-level permutation
  invariant training of deep recurrent neural networks,''
\newblock {\em IEEE/ACM Transactions on Audio, Speech, and Language
  Processing}, vol. 25, no. 10, pp. 1901--1913, Oct 2017.

\bibitem{Kinoshita18RSAN}
K.~{Kinoshita}, L.~{Drude}, M.~{Delcroix}, and T.~{Nakatani},
\newblock ``Listening to each speaker one by one with recurrent selective
  hearing networks,''
\newblock in {\em 2018 IEEE International Conference on Acoustics, Speech and
  Signal Processing (ICASSP)}, April 2018, pp. 5064--5068.

\bibitem{Roux19Phasebook}
J.~{Le Roux}, G.~Wichern, S.~Watanabe, A.~Sarroff, and J.~Hershey,
\newblock ``Phasebook and friends: Leveraging discrete representations for
  source separation,''
\newblock {\em IEEE Journal of Selected Topics in Signal Processing}, vol. PP,
  pp. 1--1, March 2019.

\bibitem{Wang2018EndtoEndSS}
Z.~Wang, J.~{Le Roux}, D.~Wang, and J.~R. Hershey,
\newblock ``End-to-end speech separation with unfolded iterative phase
  reconstruction,''
\newblock in {\em Interspeech 2018 - 19th Annual Conference of the
  International Speech Communication Association}, Sep 2018.

\bibitem{Wang19PhaseReconstruction}
Z.~{Wang}, K.~{Tan}, and D.~{Wang},
\newblock ``Deep learning based phase reconstruction for speaker separation: A
  trigonometric perspective,''
\newblock in {\em ICASSP 2019 - 2019 IEEE International Conference on
  Acoustics, Speech and Signal Processing (ICASSP)}, May 2019, pp. 71--75.

\bibitem{Luo2018TasNetSI}
Y.~Luo and N.~Mesgarani,
\newblock ``{TasNet}: Surpassing ideal time-frequency masking for speech
  separation,''
\newblock {\em CoRR}, vol. abs/1809.07454, 2018.

\bibitem{Yi19ConvTasNet}
Y.~Luo and N.~Mesgarani,
\newblock ``{Conv-TasNet}: Surpassing ideal time-frequency magnitude masking
  for speech separation,''
\newblock {\em IEEE/ACM Transactions on Audio, Speech, and Language
  Processing}, vol. PP, pp. 1--1, May 2019.

\bibitem{Shi19FurcaNet}
Z.~Shi, H.~Lin, L.~Liu, R.~Liu, and J.~Han,
\newblock ``{FurcaNet}: An end-to-end deep gated convolutional, long short-term
  memory, deep neural networks for single channel speech separation,''
\newblock {\em CoRR}, vol. abs/1902.00651, 2019.

\bibitem{Takahashi19OrPit}
N.~Takahashi, S.~Parthasaarathy, N.~Goswami, and Y.~Mitsufuji,
\newblock ``Recursive speech separation for unknown number of speakers,''
\newblock {\em CoRR}, vol. abs/1904.03065, 2019.

\bibitem{Bahmaninezhad2019TasNetSpect}
F.~Bahmaninezhad, J.~Wu, R.~Gu, S.-X. Zhang, Y.~Xu, M.~Yu, and D.~Yu,
\newblock ``A comprehensive study of speech separation: Spectrogram vs waveform
  separation,''
\newblock in {\em Interspeech 2019 - 20th Annual Conference of the
  International Speech Communication Association}, Sept. 2019, pp. 4574--4578.

\bibitem{Yin2018MultitalkerSS}
L.~Yin, Z.~Wang, R.~Xia, J.~Li, and Y.~Yan,
\newblock ``Multi-talker speech separation based on permutation invariant
  training and beamforming,''
\newblock in {\em Interspeech 2018 - 19th Annual Conference of the
  International Speech Communication Association}, Sep 2018.

\bibitem{Drude2018SpecialIssue}
L.~Drude and R.~Haeb-Umbach,
\newblock ``Integration of neural networks and probabilistic spatial models for
  acoustic blind source separation,''
\newblock {\em IEEE Journal of Selected Topics in Signal Processing}, vol. 13,
  no. 4, pp. 815--826, Aug 2019.

\bibitem{Zmolikova17SpeakerBeam}
{Katerina Zmol{\'{\i}}kov{\'{a}} and Marc Delcroix and Keisuke Kinoshita and
  Takuya Higuchi and Atsunori Ogawa and Tomohiro Nakatani},
\newblock ``Speaker-aware neural network based beamformer for speaker
  extraction in speech mixtures,''
\newblock in {\em Interspeech 2017 - 18th Annual Conference of the
  International Speech Communication Association}, Sep 2017, pp. 2655--2659.

\bibitem{Higuchi16RobustMVDR}
T.~Higuchi, N.~Ito, T.~Yoshioka, and T.~Nakatani,
\newblock ``Robust {MVDR} beamforming using time-frequency masks for
  online/offline {ASR} in noise,''
\newblock in {\em 2016 IEEE International Conference on Acoustics, Speech and
  Signal Processing (ICASSP)}, Mar 2016, pp. 5210--5214.

\bibitem{Luo2019FaSNet}
Y.~{Luo}, E.~{Ceolini}, C.~{Han}, S.-C. {Liu}, and N.~{Mesgarani},
\newblock ``{{FaSNet}: Low-latency Adaptive Beamforming for Multi-microphone
  Audio Processing},''
\newblock {\em arXiv e-prints}, p. arXiv:1909.13387, Sep 2019.

\bibitem{Wichern2019WHAM}
G.~Wichern, J.~Antognini, M.~Flynn, L.~R. Zhu, E.~McQuinn, D.~Crow, E.~Manilow,
  and J.~{Le Roux},
\newblock ``{WHAM!}: Extending speech separation to noisy environments,''
\newblock in {\em Interspeech 2019 - 20th Annual Conference of the
  International Speech Communication Association}, Sept. 2019.

\bibitem{Gu19TasNetSpatialFeatures}
R.~Gu, J.~Wu, S.~Zhang, L.~Chen, Y.~Xu, M.~Yu, D.~Su, Y.~Zou, and D.~Yu,
\newblock ``End-to-end multi-channel speech separation,''
\newblock {\em CoRR}, vol. abs/1905.06286, 2019.

\bibitem{Zhang2017GlottalMB}
Y.~Zhang, D.~A.~F. Flor{\^e}ncio, and M.~Hasegawa-Johnson,
\newblock ``Glottal model based speech beamforming for ad-hoc microphone
  arrays,''
\newblock in {\em Interspeech 2017 - 18th Annual Conference of the
  International Speech Communication Association}, Sep 2017.

\bibitem{Raffel14Mireval}
C.~Raffel, B.~Mcfee, E.~J. Humphrey, J.~Salamon, O.~Nieto, D.~Liang, D.~P.~W.
  Ellis, C.~C. Raffel, B.~Mcfee, and E.~J. Humphrey,
\newblock ``{mir\_eval}: a transparent implementation of common {MIR}
  metrics,''
\newblock in {\em Proceedings of the 15th International Society for Music
  Information Retrieval Conference, ISMIR}, 2014.

\bibitem{Povey11Kaldi}
D.~Povey, A.~Ghoshal, G.~Boulianne, L.~Burget, O.~Glembek, N.~Goel,
  .~Hannemann, P.~Motlicek, Y.~Qian, P.~Schwarz, J.~Silovsky, G.~Stemmer, and
  K.~Vesely,
\newblock ``The {Kaldi} speech recognition toolkit,''
\newblock in {\em IEEE 2011 Workshop on Automatic Speech Recognition and
  Understanding}. Dec 2011, IEEE Signal Processing Society.

\bibitem{Garofolo1993WSJ}
J.~Garofolo, D.~Graff, D.~Paul, and D.~Pallett,
\newblock ``{CSR-I} ({WSJ0}) complete {LDC93S6A},''
\newblock {\em Web Download. Philadelphia: Linguistic Data Consortium}, 1993.

\bibitem{Drude18SMSWSJ}
L.~{Drude}, J.~{Heitkaemper}, C.~{Boeddeker}, and R.~{Haeb-Umbach},
\newblock ``{SMS-WSJ: Database, performance measures, and baseline recipe for
  multi-channel source separation and recognition},''
\newblock {\em CoRR}, vol. abs/1910.13934, 2019.

\end{thebibliography}

\end{document}